\documentclass[aps,twocolumn, superscriptaddress , amsmath, amssymb, showpacs, showkeys,reprint]{revtex4-1}
\usepackage{graphicx}
\usepackage{dcolumn}
\usepackage{bm}
\usepackage{upgreek}
\usepackage{times}
\usepackage{mathrsfs}
\usepackage{multirow}
\usepackage{fancyhdr} 
\usepackage{hyperref} 
\begin{document}

\title{Unfolding the band structure of disordered solids:\\from bound states to high-mobility Kane fermions}
\author{O.~Rubel}
\email[]{rubelo@tbh.net}
\affiliation{Thunder Bay Regional Research Institute, 980 Oliver Road, Thunder Bay, Ontario P7B 6V4, Canada}
\affiliation{Department of Physics, Lakehead University, 955 Oliver Road, Thunder Bay, Ontario P7B 5E1, Canada}
\author{A. Bokhanchuk}
\affiliation{Thunder Bay Regional Research Institute, 980 Oliver Road, Thunder Bay, Ontario P7B 6V4, Canada}
\author{S. J. Ahmed}
\affiliation{Department of Materials Science and Engineering, McMaster University, Hamilton, Ontario L8S4L8, Canada}
\author{E. Assmann}
\affiliation{Institute for Solid State Physics, Vienna University of Technology, Wiedner Hauptstra{\ss}e 8-10, 1040 Vienna, Austria}

\date{\today}

\begin{abstract}
Supercells are often used in \textit{ab initio} calculations to model compound alloys, surfaces and defects. One of the main challenges of supercell electronic structure calculations is to recover the Bloch character of electronic eigenstates perturbed by disorder. Here we apply the spectral weight approach to unfolding the electronic structure of group III-V and II-VI semiconductor solid solutions. The illustrative examples include: formation of donor-like states in dilute Ga(PN) and associated enhancement of its optical activity, direct observation of the valence band anticrossing in dilute GaAs:Bi, and a topological band crossover in ternary (HgCd)Te alloy accompanied by emergence of high-mobility Kane fermions. The analysis facilitates interpretation of optical and transport characteristics of alloys that are otherwise ambiguous in traditional first-principles supercell calculations.
\end{abstract}


\maketitle

%
%
\section{Introduction}\label{Sec:Introduction}

Electronic energy eigenstates of periodic solids are traditionally represented by Bloch waves $\psi_{n,\mathbf{k}}(\mathbf{r})$ with a wave vector $\mathbf{k}$ describing the translational periodicity of a given state. Existence of the well defined wave vector in periodic structures has important consequences for their optical and transport properties imposed in the form of selection rules for inter- and intraband transitions. Therefore, the vast majority of existing theoretical studies that focus on the electronic band structure of solids use a primitive basis (or Bloch) representation.

In reality, the ideal translational periodicity of solids is disturbed by disorder in the form of structural defects, fluctuations of the chemical composition in compound alloys, magnetic disorder or even the lack of long-range order in non-crystalline solids. The disorder can significantly alter electronic structure of solids resulting in the emergence of properties that are strikingly different from their nominal constituencies \cite{Freysoldt_RMP_86_2014}. For example, electronic states associated with substitutional impurities can facilitate optical transitions in otherwise indirect semiconductors \cite{Popescu_PRL_104_2010}. Modeling of disordered structures requires construction of supercells that greatly exceed the size of a primitive basis. Since the Brillouin zone (BZ) shrinks as a result of the zone folding, the recovery of a dispersion relation $\epsilon(\mathbf{k})$ in the momentum-energy space becomes non-trivial.

The most successful approach that links the supercell band structure with the primitive basis representation is based on a Bloch spectral density \cite{Faulkner_PRB_21_1980,Durham_JPFMP_11_1981}, which is also known as a ``spectral weight" \cite{Baroni_PRL_65_1990,Dargam_PRB_56_1997,Wang_PRL_80_1998}. The spectral weight $w_n(\mathbf{k})$ amounts to a Bloch $\mathbf{k}$-character of the $n$'th energy eigenstates $\epsilon_n$ and fulfills the normalization $\sum_\mathbf{k}w_n(\mathbf{k})=1$. For symmetric structures, the Bloch wave vector $\mathbf{k}_B$ is unambiguously defined, i.e., $w_n(\mathbf{k})=\delta(\mathbf{k}-\mathbf{k}_B)$. In supercells with a broken symmetry, the energy eigenstate is characterized by broadening of $w_n(\mathbf{k})$. In the ultimate extreme of spatially localized states, the uncertainty approaches its fundamental limit of $\Delta k\sim r_\text{B}^{-1}$, where $r_\text{B}$ is the localization radius.

The spectral weight can be obtained by a Fourier transformation of local basis functions, such as atomic orbitals \cite{Dargam_PRB_56_1997,Boykin_PRB_71_2005,Lee_JPCM_25_2013}, Wannier functions \cite{Giustino_PRL_98_2007,Ku_PRL_104_2010,Konbu_JPSJ_80_2011,Berlijn_PRL_106_2011,Berlijn_PRL_108_2012,Berlijn_PRL_109_2012,Konbu_SSC_152_2012,Berlijn_PRB_89_2014} or projected local orbitals  \cite{Haverkort_arXiv_1109.4036}. In the case of a non-local basis set, such as plane waves, the spectral weight can be constructed from the Fourier expansion coefficients by gathering them in groups associated with a particular Bloch wave vector \cite{Wang_PRL_80_1998,Popescu_PRL_104_2010,Popescu_PRB_85_2012,Allen_PRB_87_2013,Medeiros_PRB_89_2014}. The latter approach is the most straightforward for implementation in solid-state \emph{ab initio} electronic structure codes, since the plane wave (PW) expansion coefficients are readily available in pseudopotential or full-potential packages.

Here we apply the ``spectral weight" approach to unfolding of the electronic structure of group III-V and II-VI semiconductor solid solutions obtained in the framework of density functional theory with the all-electron \texttt{Wien2k} package \cite{Blaha_2001}.  For illustration, we investigate (i) formation of donor-like states in dilute GaP:N and associated enhancement of its optical activity, (ii) direct observation of the valence band anticrossing in dilute GaAs:Bi, and (iii) a topological band crossover in a bulk (HgCd)Te compound alloy and related formation of high-mobility Kane fermions. These features, which would have been difficult to identify in standard band structure calculations, become apparent in the unfolded band structures produced by our method.

%
%
\section{Unfolding procedure}\label{Sec:Method}

\begin{figure*}
	\includegraphics[width=0.9\textwidth]{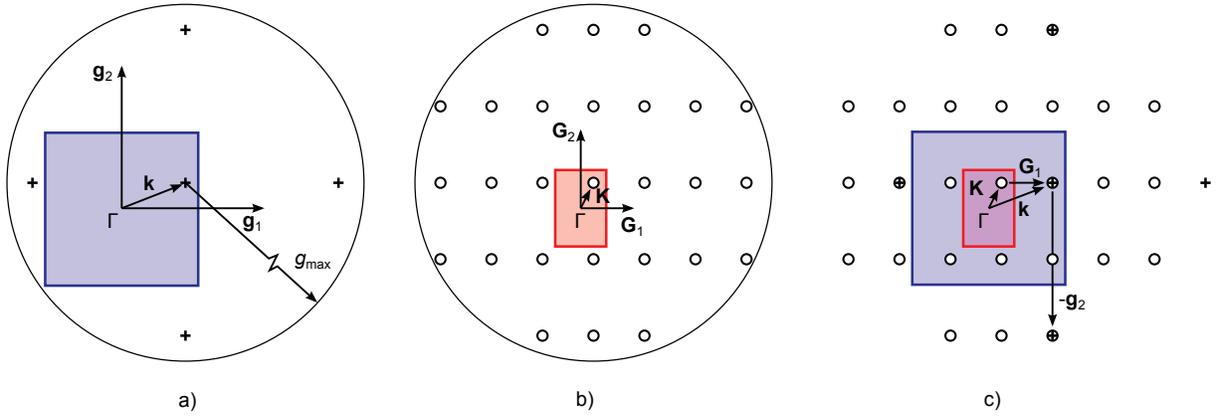}\\
	\caption{Relation between $k$-mesh in the reciprocal space for a square primitive lattice (a) and that for a $3\times2$ supercell (b). The first BZ is outlined on both panels. The overlay of both patterns is shown on the panel (c). The overlapped points form a subgroup of $\mathbf{K}$ that represents the Bloch wave vector $\mathbf{k}$. There is a total of 6 subgroups ($3\times2$), which indicates that each $K$-point in the supercell bears information about 6 $k$-points of the primitive basis.}\label{Fig:A}
\end{figure*}

Our approach to the calculation of the spectral weight is based on remapping the supercell reciprocal space with a mesh that is compatible with the translational symmetry of a primitive cell \cite{Popescu_PRL_104_2010,Popescu_PRB_85_2012,Chen_NL__2014}. Here we briefly review the basics of this method.

The PW expansion alone
\begin{equation}\label{Eq:A}
    \Psi_{n,\mathbf{K}}(\mathbf{r}) = \sum_\mathbf{G} C_{n,\mathbf{K}}(\mathbf{G}) \, \mathrm{e}^{i(\mathbf{K}+\mathbf{G})\cdot\mathbf{r}}
\end{equation}
or its combination with a local basis set (such as augmented plane-waves) is a popular choice for representing wave functions in periodic solids. Here $n$ refers to a particular eigenstate (band index), $\mathbf{K}$ is the wave vector within the first BZ and $C$ are expansion coefficients. The summation runs over a set of $K$-points repeated with periodicity of the  reciprocal lattice vectors $\mathbf{G}_1$, $\mathbf{G}_2$ and $\mathbf{G}_3$. The plane wave cut-off $G_\text{max}$ determines the range of the summation and, therefore, the completeness of the basis set. The general form of expansion (\ref{Eq:A}) is identical irrespective of whether a supercell or a primitive cell basis is used. We will employ upper-case and lower-case notations in order to distinguish between these two cases, respectively.

Figure~\ref{Fig:A}~(a,b) illustrates the reciprocal space mesh in two dimensions for a primitive cubic lattice and its supercell of the size $3\times2$. Each point on this mesh is associated with an individual PW, and can be assigned a relative ``weight" of $|C_{n,\mathbf{K}}(\mathbf{G})|^2$. When the two meshes corresponding to the primitive cell and supercell overlay as shown in Fig.~\ref{Fig:A}(c), it is possible to match the supercell $\mathbf{K}$ and the primitive Bloch wave $\mathbf{k}$ PW expansion coefficients
\begin{equation}\label{Eq:B}
    C_{n,\mathbf{K}}(\mathbf{G}) \rightarrow c_{n,\mathbf{k}}(\mathbf{g})
\end{equation}
at the points which fulfill
\begin{equation}\label{Eq:F}
    \mathbf{K} + \mathbf{G} = \mathbf{k} + \mathbf{g}~.
\end{equation}

As can be seen in Fig.~\ref{Fig:A}(c), any $K$-point transforms into $N_1 N_2 N_3$ $k$-points in the first primitive BZ under the translation
\begin{equation}\label{Eq:C}
    \mathbf{k} = \mathbf{K} + m_1\mathbf{G}_1 + m_2\mathbf{G}_2 + m_3\mathbf{G}_3
\end{equation}
with $m_i = 0,1,\ldots N_i-1$ that extends up to the scaling factor $N_i$ used when constructing the supercell along $i$'s axis. This generates a multitude of ``unfolded" Bloch wave vectors, each with its own subgroup of the PW expansion coefficients $C_{n,\mathbf{K}}(\mathbf{k}+\mathbf{g})$. Thus, the individual ``weights" of unfolded $k$-points are expressed in terms of the PW coefficients which belong to the subgroup of $\mathbf{k}$
\begin{equation}\label{Eq:E}
    w_n(\mathbf{k}) = \sum_{\mathbf{g}} |C_{n,\mathbf{K}}(\mathbf{k}+\mathbf{g})|^2~.
\end{equation}
Note that the subgroups are formed by the translation vectors $\mathbf{g}$, not $\mathbf{G}$. In order to facilitate the mapping, the supercell needs to be generated by translation of the \textit{primitive cell} along its lattice vectors in real space, which implies a simple relation between the reciprocal lattice vectors $\mathbf{g}_i=m_i\mathbf{G}_i$. If latter is not the case, an additional coordinate transformation ($m_i\mathbf{G}_i \rightarrow \mathbf{g}_i$) is required for the resultant wave vectors $\mathbf{k}$ given by Eq.~(\ref{Eq:C}).

%
%
\section{Applications}\label{Sec:Results}

\subsection{Dilute GaP:N}\label{Sec:Results:GaNP}

Group III-V dilute nitride semiconductors continue to be in the focus since the 1990's as a material system for long-wavelength telecommunication and photovoltaic applications \cite{Kondow_JCG_164_1996,Kondow_JJAP_35_1996}. In spite of the fact that nitrides (GaN and AlN) are wide-bandgap semiconductors, addition of a small fraction of nitrogen (a few percent) in the host III-V semiconductors, e.g. GaAs, results in a drastic reduction of their energy gap. This narrowing of the band gap is attributed to an anticrossing between extended states of the host conduction band and the localized nitrogen resonant states \cite{Shan_PRL_82_1999}.

\begin{figure*}
	\includegraphics[width=0.9\textwidth]{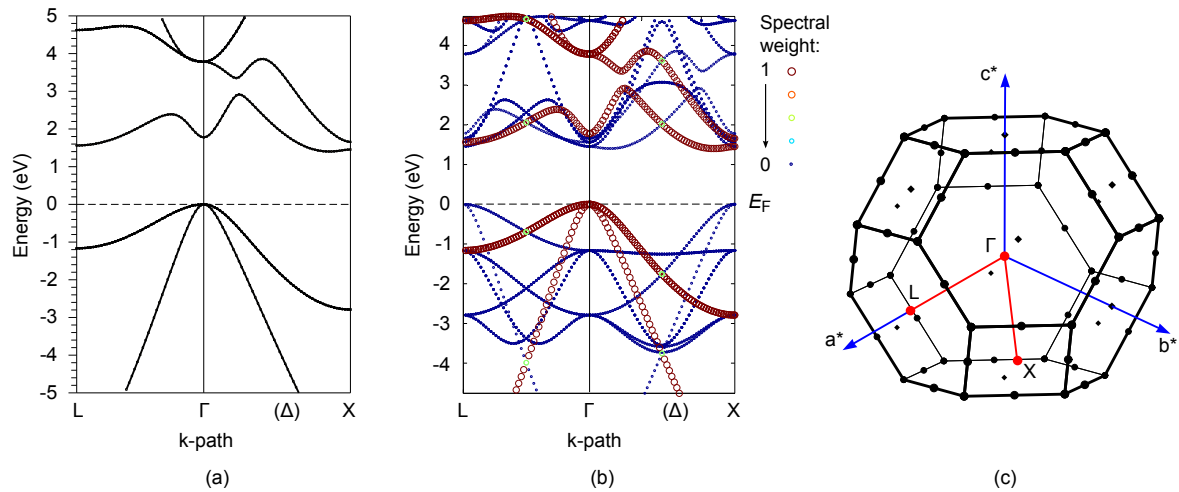}\\
	\caption{Band structure of GaP: 2-atoms primitive unit cell (a) and 16-atoms $2\times2\times2$ supercell (b). The supercell band structure is ``obfuscated" due to the zone folding. Fat bands on panel (b) illustrate the unfolded band structure where the symbols size and color represent the Bloch spectral weight. Both cells have the Brillouin zone shown on panel (c). Here and in all figures hereafter the Fermi energy is set to zero.}\label{Fig:B}
\end{figure*}

\begin{figure*}
	\includegraphics[width=0.9\textwidth]{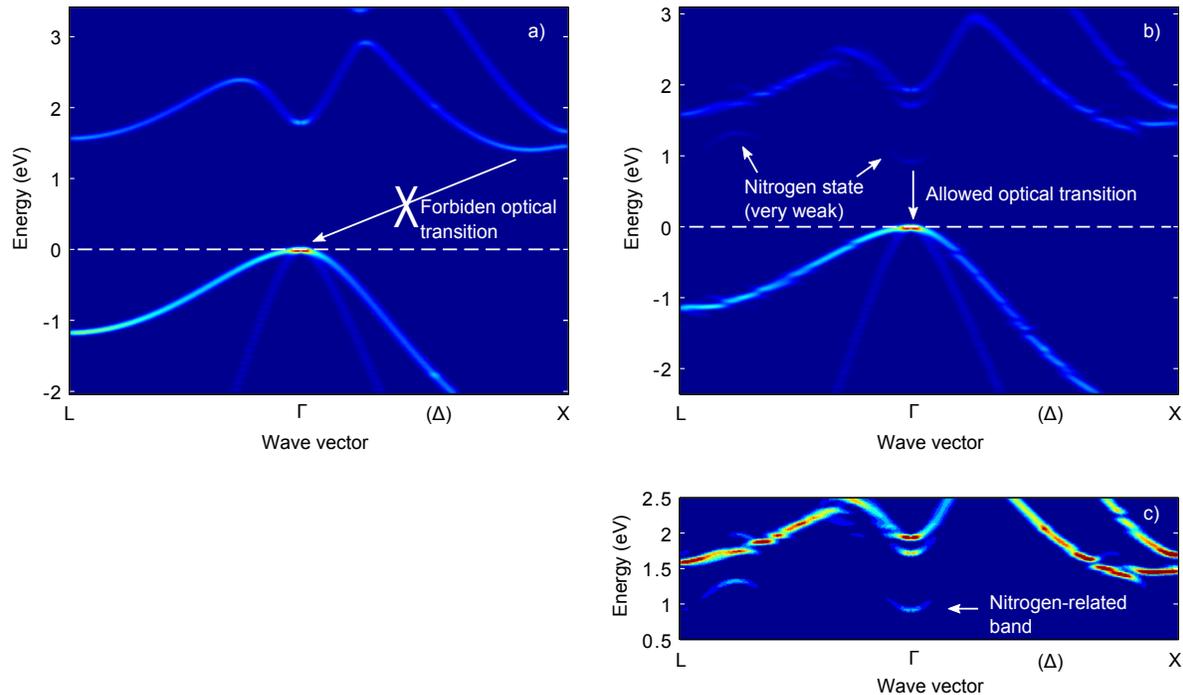}\\
	\caption{Band structure of Ga$_{64}$P$_{64}$ (a) and Ga$_{64}$N$_{1}$P$_{63}$ (b,~c) supercell unfolded to the primitive Bloch representation. A very weak signature of a nitrogen-related conduction band can be seen within the GaP energy gap (b). Panel (c) shows a snapshot of Ga$_{64}$N$_{1}$P$_{63}$ conduction band with the contrast artificially enhanced by using a non-linear (square root) intensity scale. The nitrogen band introduces a direct optical transition (b) thereby stimulating radiative optical transitions in Ga(NP) \cite{Groves_US3725749_1973}.}\label{Fig:C}
\end{figure*}

Ga(NP) was a progenitor of modern dilute nitrides \cite{Thomas_PR_150_1966}. It was shown that nitrogen and its complexes behave in GaP as isoelectronic traps by creating a tail of localized states in the vicinity of the conduction band edge \cite{Kent_PRB_64_2001}. The nitrogen-related states facilitate radiative recombination of optical excitations, which is otherwise suppressed due to the indirect band structure of GaP \cite{Groves_US3725749_1973}. However, it should be noted that disorder also introduces non-radiative channels that have  an adverse effect on the internal quantum efficiency of light-emitting devices \cite{Buyanova2004,Rubel_PRB_73_2006}. These are the main features of Ga(NP), which will be used as a benchmark in order to prove the validity of the proposed method.

The band structure of GaP is presented in Fig.~\ref{Fig:B}(a). (Computational details are provided in the Appendix.) The calculated band structure reproduces experimental features of the GaP band structure, namely, the indirect band gap with the conduction band minimum along $\Delta$-path near X-point followed by L and $\Gamma$ valleys in the direction of increasing the electron energy (see Ref.~\onlinecite{Adachi1999}, pp. 198-199). Our calculations do not purport to reproduce an experimental value of the band gap in GaP. The calculated energy gap of GaP is only about 60\% of its experimental value, which is due to a shortcoming of the density functional theory (DFT) in its Kohn-Sham formulation \cite{Kohn_PR_140_1965}. This drawback is not critical for the purpose of our study since qualitative prediction of the optical emission spectrum is beyond the scope of our work. Next, we repeat the band structure calculation for a 16-atoms supercell of GaP. Results are presented in Fig.~\ref{Fig:B}(b), which exemplifies a zone folding that hinders analysis of the band structure of supercells. Even for such a comparatively small supercell ($2\times2\times2$), the direct or indirect character of the band gap is obscured.

Now we apply the procedure described in Sec.~\ref{Sec:Method} in the attempt to recover the GaP band structure in its conventional Bloch representation from a 128-atom supercell. The unfolded band structure is shown in Fig.~\ref{Fig:C}(a) and can be directly compared to that in Fig.~\ref{Fig:B}(a), including the indirect band gap and ordering of valleys in the conduction band. It is important to note that all points are sharp within the limit of a Gaussian smearing applied. This feature is inherent to Bloch states with a well defined wave vector $\mathbf{k}$, as anticipated for a structure in the absence of a disorder. The brightness of the spots is determined by the magnitude of the corresponding Bloch character as well as by the degeneracy. The distinct brightness of the valence and conduction bands in Fig.~\ref{Fig:C}(a) is due to a degeneracy of the corresponding eigenvalues. This band structure will be used as a reference when studying effects of disorder.

The symmetry of GaP supercell is then disturbed by introducing nitrogen as an isoelectronic substitutional impurity. The unfolded band structure of Ga$_{64}$N$_{1}$P$_{63}$ is shown in Fig.~\ref{Fig:C}(b). Nitrogen incorporation results in the emergence of a new band beneath the host conduction band of GaP. At the same time, the valence band remains almost unperturbed. The nitrogen-related states are better seen in Fig.~\ref{Fig:C}(c) where the conduction band region is shown with enhanced contrast. Contrary to the extended states of GaP, the nitrogen states do not have a well defined $\mathbf{k}$ and they are only weakly present at the $\Gamma$-point. These results are consistent with the localized nature of nitrogen states in GaP and confirm their role as recombination centers for optical excitations, albeit slow and inefficient \cite{Kent_PRB_64_2001,Niebling_JPCM_20_2008}.

The intrinsic limitations of dilute Ga(NP) stimulated exploration of alternative materials for room-temperature optical emitters. Ga(NAsP) is one of the recent developments in the family of dilute nitrides that holds promise for the realization of monolithic optoelectronic integrated circuits on silicon substrates \cite{Kunert_APL_88_2006,Yu_APL_88_2006}.

It is important to acknowledge that study of random alloys using small supercells without proper configurational averaging may result in artifacts observed in the unfolded band structure due to a periodic ordering \cite{Haverkort_arXiv_1109.4036}. This indicates a need to test convergence of the resultant band structure with respect to the size of the supercell as well as the number of random configurations. At this point, the simulation scale may become computationally prohibitive for DFT. Development of an effective Hamiltonian, which incorporates effects of compositional disorder, based on first-principles Wannier functions \cite{Berlijn_PRL_106_2011} is an efficient technique that enables to overcome this limitation. Furthermore, the configurational averaging can be expedited by constructing special quasirandom structures with relevant radial correlation functions tailored to match a perfectly random structure \cite{Zunger_PRL_65_1990}.

\begin{figure*}
	\includegraphics[width=0.9\textwidth]{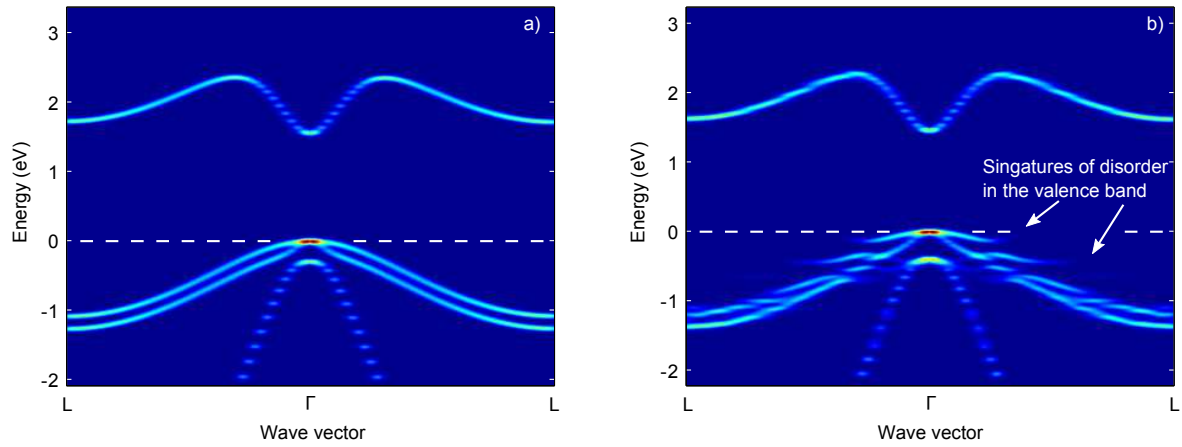}\\
	\caption{Band structure of Ga$_{64}$As$_{64}$ (a) and Ga$_{64}$As$_{63}$Bi$_{1}$ (b) supercells unfolded to the primitive Bloch representation. Bismuth incorporation leads to perturbations in the valence band, reduction of the band gap and enhanced spin-orbit splitting.}\label{Fig:D}
\end{figure*}

\subsection{Dilute GaAs:Bi}\label{Sec:Results:GaAsBi}

The above example of Ga(NP) shows that incorporation of nitrogen is responsible for the modulation of the conduction band in the host semiconductor. Alloying of GaAs with bismuth has been proposed as a complementary approach for engineering of the valence band \cite{Francoeur_APL_82_2003}. Reduction of the band gap in Ga(AsBi) is explained using a semi-empirical valence band anticrossing model \cite{Alberi_APL_91_2007}. The model is similar in spirit to the band anticrossing of dilute nitrides \cite{Shan_PRL_82_1999}. Recent electronic structure calculations of GaAs$_{1-x}$Bi$_x$ \cite{Deng_PRB_82_2010} performed beyond the popular band anticrossing model provide important insight to the structure of the valence band. However, the valuable link between electronic eigenstates of the alloy and their dispersion characteristics is overlooked in this analysis. This relation can be recovered using the unfolding procedure.

The band structures of GaAs and Ga$_{64}$As$_{63}$Bi$_{1}$ are shown in Fig.~\ref{Fig:D}. Their comparison allows us to draw the following conclusions: (i) incorporation of Bi  in GaAs mostly affects the valence band, (ii) no bound states are found above the valence band edge of the host GaAs, (iii) the energy gap shrinks along with enhancement of the spin-orbit splitting. These features are consistent with the experimental trends \cite{Alberi_APL_91_2007} and prediction from first-principles of bismuth-related signatures of localization \textit{beneath} the host valence band maximum \cite{Deng_PRB_82_2010}.

Finally, we would like to emphasize a fundamental difference between GaAs$_{1-x}$Bi$_x$ and the majority of other group III-V ternary semiconductors, including GaAs$_{1-x}$Sb$_x$ alloy. GaBi is a metal with an anomalous order of bands \cite{Janotti_PRB_65_2002}, which is reminiscent of the better known $\alpha$-Sn with a band inversion or negative band gap caused by spin-orbit effects \cite{Kufner_N_24_2013}. Thus, the electronic structures of GaAs and GaBi have different topology, which is not the case in GaAs-GaSb, for instance. The electronic structure of GaAs$_{1-x}$Bi$_x$ cannot evolve ``smoothly" in the range $0<x<1$ and must undergo a topological phase transition. This transition is accompanied by a gradual transformation of the host GaAs parabolic conduction band to a graphene-like cone with increasing $x$; electrons ultimately become massless fermions as it will be shown later. Verification of this prediction would require electron transport measurements in GaAs$_{1-x}$Bi$_x$ for a wide range of $x$. So far, the successful incorporation of Bi in GaAs under 12\% has been reported \cite{Batool_chapter_GaAsBi_2013}. The lattice mismatch of 12\% between GaAs and GaBi is one of the main factors that limits their solubility.

\subsection{(HgCd)Te alloy}\label{Sec:Results:HCT}

\begin{figure*}
	\includegraphics[width=0.9\textwidth]{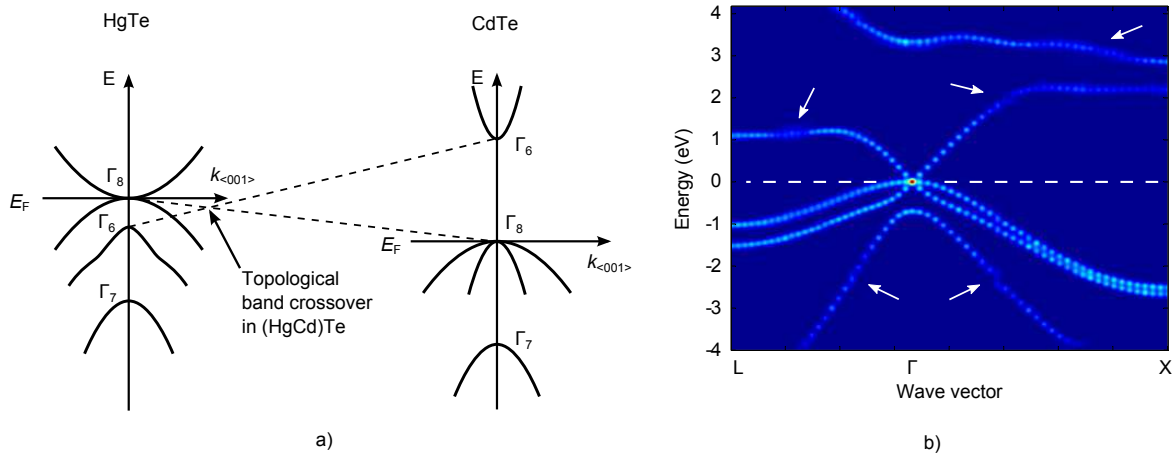}\\
	\caption{Band order and symmetry labels for HgTe and CdTe compounds (a). These two compounds exhibit a band inversion ($\Gamma_6 \leftrightarrow \Gamma_8$) and, thus, have different ``topology". As the composition Hg$_{1-x}$Cd$_x$Te varies between two binary compounds, a topological band crossover occurs. The crossover takes place near $x_\text{c}\approx0.19$ and is accompanied by emergence of massless Kane fermions at the $\Gamma$-point (b). Regions of the band structure perturbed by disorder are labeled with arrows. The disorder affects only the electronic states located $1-3$~eV above and below the Fermi energy.}\label{Fig:E}
\end{figure*}

Hg$_{1-x}$Cd$_x$Te (HCT) is an example of a material system with the topological band inversion \cite{Hasan_RMP_82_2010}, which is similar to GaAs$_{1-x}$Bi$_x$ discussed in the preceding Sec.~\ref{Sec:Results:GaAsBi}. The arrangement of bands in binary HgTe and CdTe as well as their symmetry are shown at Fig.~\ref{Fig:E}(a). The conduction band minimum and the valence band maximum of CdTe have $\Gamma_6$ and $\Gamma_8$ symmetries, respectively. In HgTe, the order of bands is inverted due to a strong spin-orbit interaction \cite{Zawadzki_AP_23_1974}. The crossover between the $\Gamma_6$ and $\Gamma_8$ bands is inevitable in the course of a gradual change in the composition of ternary Hg$_{1-x}$Cd$_x$Te alloy as illustrated by dashed lines in Fig.~\ref{Fig:E}(a). The prominent feature of HCT is the presence of massless Kane fermions \cite{Kane_JPCS_1_1957} at the crossover composition  (Fig.~\ref{Fig:E}b), whose experimental observation was recently reported by \citet{Orlita_NP_10_2014}.

\begin{figure*}
	\includegraphics[width=\textwidth]{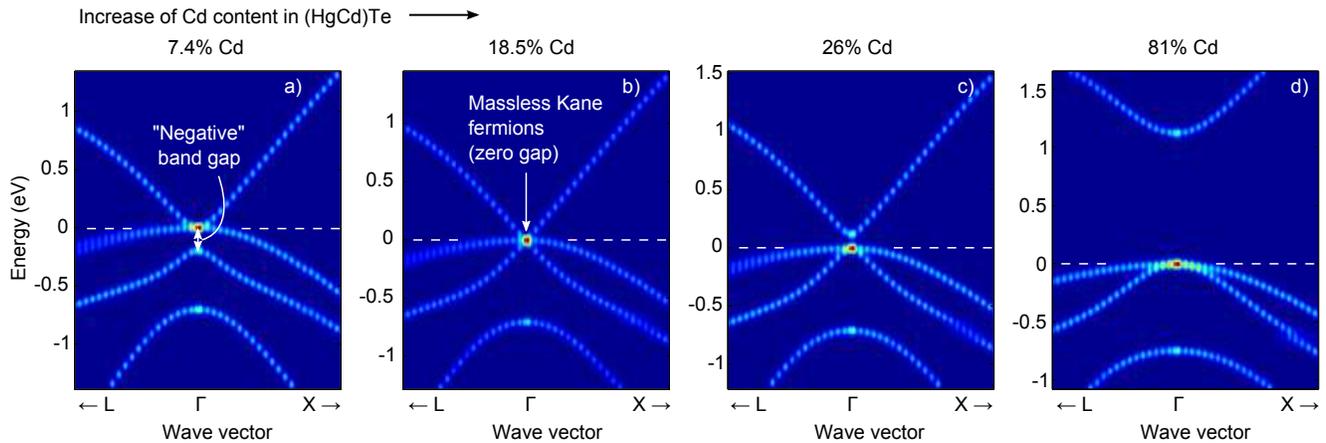}\\
	\caption{Evolution of the band structure in ternary (HgCd)Te alloy near the $\Gamma$-point as a function of the chemical composition: Hg$_{25}$Cd$_{2}$Te$_{27}$~(a), Hg$_{22}$Cd$_{5}$Te$_{27}$ (b), Hg$_{20}$Cd$_{7}$Te$_{27}$ (c), Hg$_{5}$Cd$_{22}$Te$_{27}$ (d). The transition from a semimetal (a) to an insulator (c,d) occurs by passing through a Kane point (b). The nearly linear dispersion $E\propto|\mathbf{k}|$, which is characteristic of the Kane point, persists in the conduction band after opening of the band gap (c).}\label{Fig:F}
\end{figure*}

The evolution of the Hg$_{1-x}$Cd$_x$Te band structure as a function of composition is shown in Fig.~\ref{Fig:F}. The composition range was chosen to cover the transition from a semimetal with a negative band gap to an insulator. The negative gap gradually shrinks with increasing the Cd content (Fig.~\ref{Fig:F}a) until three-fold degeneracy is established at a critical composition (Fig.~\ref{Fig:F}b). At this composition, the light hole and electron masses vanish near $\Gamma$ as it is evident from the conical shape of their dispersion. \citet{Orlita_NP_10_2014} stressed that Kane fermions are not protected by symmetry, unlike Dirac fermions. The emergence of Kane fermions corresponds to a critical chemical composition $x_\text{c}$. The critical composition is sensitive to extrinsic factors, such as  temperature, pressure.

The band structure calculations yield the critical cadmium content of $x_\text{c}\approx0.19\pm0.04$ \textit{vs}. $0.15\ldots0.17$ observed experimentally \cite{Rogalski_RPP_68_2005}. The modest level of discrepancy is largely due to success of the Tran-Blaha modified Becke and Johnson (mBJ) exchange potential \cite{Tran_PRL_102_2009} in correcting the energy gap error introduced in regular LDA (local density approximation) calculations. The LDA-mBJ values of the band gap in binary CdTe and HgTe are 1.56 and $-0.25$~eV, respectively, compared to their experimental values of 1.65 and $-0.3$~eV \cite{Rogalski_RPP_68_2005}. It is interesting that LDA-mBJ accurately reproduces not only the energy gap for insulators, but also performs well for semimetals. Without mBJ potential, the LDA results for the band gap in CdTe and HgTe are 0.3 and $-0.8$~eV. The Kane fermions can still be observed, but the critical concentration is heavily shifted towards Cd-rich composition $x_\text{c}\approx0.8$.

Further increase of the cadmium content beyond $x_\text{c}$ leads to a narrow-gap semiconductor with a highly non-parabolic conduction band (compare Figs.~\ref{Fig:F}c and \ref{Fig:F}d). Apparently, there is no ambiguity in the Bloch character for all states near the Fermi energy irrespective of the Hg$_{1-x}$Cd$_x$Te composition (Fig.~\ref{Fig:F}). This result indicates that charge transport characteristics of HCT do not degrade as a result of the alloy scattering as dramatically as in dilute nitride semiconductors \cite{Fahy_APL_83_2003}. Our results explain the previously established experimental facts for HCT, such as the exceptional electron mobility exceeding $10^5$~cm$^2$V$^{-1}$s$^{-1}$ at low temperature \cite{Edwall_JEM_26_1997} with its maximum value of $\sim10^6$~cm$^2$V$^{-1}$s$^{-1}$ at the composition that corresponds the topological band crossover \cite{Yoo_JAP_81_1997}.

There are two arguments why the disorder has such a mild effect on the band structure of Hg$_{1-x}$Cd$_x$Te. First, Hg and Cd have almost the identical atomic radii that results in no local lattice distortions when one element is interchanged by the other. Second, both elements have almost identical the energy level of their valence $s$ and $p$ electrons which leads to comparable electronegativity of the two elements. None of these conditions apply to Ga(PN) or Ga(AsBi) alloys. As a result, their band structure is heavily perturbed by the disorder.

%
%
\section{Conclusions}\label{Sec:Conclusions}

Effects of alloying on the electronic structure of Ga(NP), Ga(AsBi) and (HgCd)Te were studied from first principles. Particular emphasis was placed on the Bloch character of the energy bands, which is evaluated using a Bloch spectral function technique specially tailored to the density functional full-potential package \texttt{Wien2k}. The success of the theory in predicting chemical trends is validated by reproducing well-known properties of dilute GaP:N. The calculations yield a nitrogen-related band near the bottom of the host GaP conduction band. A simultaneous enhancement of the optical activity observed in GaP:N is attributed to a weak $\Gamma$-character of nitrogen-related states within the band gap of GaP.

In contrast to the role of nitrogen, incorporation of bismuth in GaAs leads to perturbations in the valence band without any noticeable degradation of the conduction band. The valence band dispersion in the energy range of $0\ldots0.6$~eV below the Fermi energy is significantly affected by disorder. Uncertainties in the Bloch character of those states indicate the lost approximate translational symmetry of the host GaAs. Based on the analogy between GaAs-GaBi and CdTe-HgTe material systems, it is anticipated that the electron effective mass will be reduced as the bismuth content grows. This trend is common to all compound alloys that combine a semiconductor and a semimetal with the topological band inversion.

A topological band crossover was illustrated by the example of a bulk ternary (HgCd)Te alloy, which manifests in the formation of high-mobility Kane fermions previously reported experimentally. The massless dispersion develops at the composition that corresponds to the semimetal-insulator transition. In contrary to Ga(NP) and Ga(AsBi), the compositional disorder practically does not disturb the bottom of conduction band and the top of valence band in (HgCd)Te, which translates to its exceptional charge transport characteristics.

\begin{table*}
    \caption{Calculation parameters for the compounds studied.}\label{Table:A}
    \begin{ruledtabular}
        \begin{tabular}{l c c c}
            Parameters  & GaP:N  & GaAs:Bi & Hg$_{27-x}$Cd$_{x}$Te$_{27}$ \\
            \hline
            $a_0$ (\AA) & 5.41 & 5.61 & 6.41\footnote{The binary average value of $a_0$ is used throughout the calculations due to a minor difference (less that 1\%) in the unit cell volume between HgTe and CdTe.}\\
            Supercell size & $4\times4\times4$ & $4\times4\times4$ & $3\times3\times3$ \\
            Number of atoms & 128 & 128 & 54 \\
            \multirow{3}{*}{Muffin tin radii $R^\text{MT}$ (Bohr)} & 1.93 (Ga) & 2.17 (Ga) & 2.50 (Hg) \\
             & 1.93 (P) & 2.06 (As) & 2.50 (Cd) \\
             & 1.60 (N) & 2.28 (Bi) & 2.49 (Te) \\
             \multirow{3}{*}{Valence electrons} & 3d$^{10}$4s$^2$4p$^1$ (Ga) & 3d$^{10}$4s$^2$4p$^1$ (Ga) & 5p$^{6}$5d$^10$6s$^2$ (Hg) \\
             &  3s$^2$3p$^3$ (P) & 3d$^{10}$4s$^2$4p$^3$ (As) & 4p$^{6}$4d$^{10}$5s$^2$ (Cd) \\
             & 2s$^2$2p$^3$ (N) & 5d$^{10}$6s$^2$6p$^3$ (Bi) & 4d$^{10}$5s$^2$5p$^4$ (Te) \\
            Optimization of atomic positions & yes & yes & no\footnote{Forces did not exceed 2~mRy/Bohr.} \\
            Reciprocal $k$-mesh & $2\times2\times2$ & $2\times2\times2$ & $3\times3\times3$ \\
            $R^\text{MT}_\text{min}K_\text{max}$ & 7 & 7 & 6 \\
            Spin-orbit coupling & no & yes & yes \\
            Direction of magnetization & $\cdots$ & [001] & [001] \\
            mBJ potential & no & yes & yes \\
        \end{tabular}
    \end{ruledtabular}
\end{table*}

%
%
\begin{acknowledgments}
Authors are indebted to Profs.~Marek Niewczas and Peter~Blaha for stimulating discussions and critical reading of the manuscript. OR and SJA would like to acknowledge funding provided by the Natural Sciences and Engineering Research Council of Canada under the Discovery Grant Program 386018-2010; EA acknowledges the support from a ``Vienna University of Technology innovative project grant".
\end{acknowledgments}

%
%
\appendix*\section{Computational details}

The first-principles calculations were carried out using density functional theory and the linear augmented plane wave method implemented in the \texttt{Wien2k} package \cite{Blaha_2001}. The local density approximation \cite{Perdew_PRB_45_1992} has been used for the exchange correlation functional. The Tran-Blaha modified Becke and Johnson (mBJ) potential \cite{Tran_PRL_102_2009} was applied to GaAs:Bi and (HgCd)Te compounds in order to improve their band gaps. Sampling of the Brillouin zone, muffin tin radii $R^\text{MT}$, the product $R^\text{MT}_\text{min}K_\text{max}$, which determines the accuracy of a plane wave expansion of the wave function, and other parameters are summarized in Table~\ref{Table:A}.

The supercells were built on the basis of primitive cells instead of conventional ones. The self-consistent lattice constants $a_0$ of binary hosts were used in the calculations (Table~\ref{Table:A}). Where indicated, the internal degrees of freedom were relaxed by minimizing Hellmann-Feynman forces acting on atoms below 2~mRy/Bohr.
Calculations of the spectral weight were performed using \texttt{fold2Bloch} package, which is available from \texttt{GitHub}.

%
%

\end{document}